\documentclass[showkeys,showpacs]{revtex4}
\usepackage{times}
\usepackage{epsfig,amssymb,amsfonts,amsmath}
\usepackage[usenames]{color}

\usepackage{amssymb}
\usepackage{amsmath}
\usepackage{amsfonts}
\usepackage{graphicx}
\usepackage[usenames]{xcolor}
\usepackage{ulem} 

\usepackage{cancel}
\usepackage{extarrows}
\providecommand{\av}[1]{\left\langle #1 \right\rangle} %

\begin{document}

\title[ ]{Reply to ``Rescuing the MaxEnt treatment for $q$-generalized entropies'' by A. Plastino and M. C. Rocca}
\author{$^{1}$Thomas Oikonomou}
\email{thomas.oikonomou@nu.edu.kz}
\author{$^{2}$G. Baris Bagci}
\email{gbb0002@hotmail.com}

\affiliation{$^{1}$Department of Physics, School of Science and Technology, Nazarbayev University, 53 Kabanbay Batyr Ave., Astana 010000, Kazakhstan}
\affiliation{$^{2}$Department of Materials Science and Nanotechnology Engineering, TOBB University of Economics and Technology, 06560 Ankara, Turkey}

\begin{abstract}
Plastino and Rocca [Physica A 491, 1023 (2018)] recently criticized our work [Phys. Lett. A 381, 207 (2017)] on the ground that one should use functional calculus instead of the ordinary calculus adopted by us in the entropy maximization procedure. We simply point out that our work requires right from the beginning $\frac{\partial S_{q}}{\partial U} = \beta$, whereas the formalism of Plastino and Rocca yields $\frac{\partial S_{q}}{\partial U} =q \beta Z^{1-q}\neq\beta$. Therefore, the work of Plastino and Rocca is irrelevant for our work.

\end{abstract}

\eid{ }
\date{\today }
\startpage{1}
\endpage{1}
\maketitle


In Ref. \cite{PlastinoRocca2017}, considering the Tsallis entropy $S_q$, Plastino and Rocca obtains the following equation (see Eq. (2.8) in Ref. \cite{PlastinoRocca2017}) 
\begin{eqnarray}\label{eq01}
\frac{q}{1-q} P_{i}^{q-1} + \lambda_{1} U_{i} + \lambda_{2} = 0 .
\end{eqnarray}
Multiplying the above equation with $P_{i}$, summing over the index $i$ and then using $\sum_{i} P_{i}^{q} = (1-q) S_{q}+1$ by the definition of Tsallis entropy, we obtain
\begin{eqnarray}\label{eq02}
S_{q} = \beta U Z^{1-q}+\frac{Z^{1-q}-1}{1-q} \,,
\end{eqnarray}
where $U$ is the average internal energy (which is denoted as $\langle U \rangle$ in Ref. \cite{PlastinoRocca2017}), $Z$ is the partition function. Note that we have also used Eqs. (2.10) and (2.11) given in Ref. \cite{PlastinoRocca2017} in order to obtain Eq. (2) above \cite{note1}. Then, from Eq. (2), after some simple algebra, we see that
\begin{eqnarray}\label{eq03}
\frac{\partial S_{q}}{\partial U} = q\beta Z^{1-q}\,. 
\end{eqnarray}
However, one can check that right from the beginning (see Eq. (1) in Ref. \cite{OikBagci2017}), we have assumed that $\frac{\partial S_{q}}{\partial U} = \beta$ holds in Ref. \cite{OikBagci2017}. For example, Eq. (2) in Ref. \cite{OikBagci2017}, which is pivotal for the subsequent analysis, cannot hold if $\partial S_q/\partial U\neq\beta$. Whether the right hand side of the expression $\frac{\partial S_{q}}{\partial U}$ depends only on $\beta$ or not changes all the mathematical and physical structure independent of the calculus (ordinary or functional) one uses.  Accordingly, Eq. (\ref{eq03}) above reveals that the work by Plastino and Rocca is incommensurable with the theoretical frame of Ref. \cite{OikBagci2017}.

Concerning the part of Ref. \cite{PlastinoRocca2017} related to the R\'enyi entropy, we note that the distribution adopted in Ref. \cite{PlastinoRocca2017} is not of the same form studied by us in Ref. \cite{OikBagci2017} (compare Eq. (28) in \cite{OikBagci2017} with no explicit $\av{U}$ appearance in the probability distribution and Eq. (3.14) in \cite{PlastinoRocca2017} for the explicit appearance of the internal energy $\av{U}$). However, it is interesting to note that it is in fact Plastino and Rocca (together with F. Pennini) who showed that the distribution in Eq. (3.14) in Ref. \cite{PlastinoRocca2017} leads to the fact that there can be no consistent thermodynamics if one adopts this particular distribution (see Sec. IV in Ref. \cite{PRP} for this issue).

To sum up, Ref. \cite{PlastinoRocca2017} has no direct implication for our work \cite{OikBagci2017} considering the Tsallis $q$-entropy, since the former work leads to $\frac{\partial S_{q}}{\partial U} = q\beta Z^{1-q}$ while our work assumes $\frac{\partial S_{q}}{\partial U} = \beta$ (and only functions in this particular context) right from the beginning. Considering the part related to the R\'enyi entropy, we note that the distribution used in Ref. \cite{PlastinoRocca2017} is different than ours in Ref. \cite{OikBagci2017}.

\begin{acknowledgments}
This research is partly supported by state-targeted program ``Center of Excellence for Fundamental and Applied Physics" (BR05236454) by the Ministry of Education and Science of the Republic of Kazakhstan and ORAU grant entitled ``Casimir light as a probe of vacuum fluctuation simplification" with PN 17098.
\end{acknowledgments}



\begin{thebibliography}{20}

\bibitem{PlastinoRocca2017} A. Plastino \& M. C. Rocca, Physica A \textbf{491} (2018) 1023.



\bibitem{note1} Note that there is also a typo in Eq. (2.10) in Ref. \cite{PlastinoRocca2017}, which should instead read as $\lambda_{1} = - q\beta Z^{1-q}$.  



\bibitem{OikBagci2017} T. Oikonomou \& G.B. Bagci, Phys. Lett. A \textbf{381} (2017) 207.



  

\bibitem{PRP} A. Plastino, M. C. Rocca and F. Pennini, Phys. Rev. E \textbf{94} (2016) 012145.


























































































  
































\end{thebibliography}
\end{document}